\RequirePackage{lineno}
\documentclass[aps,twocolumn,showpacs,byrevtex,prl,reprint]{revtex4-1}

\usepackage{graphicx}
\usepackage{dcolumn}
\usepackage{bm}
\usepackage{rotating}
\usepackage{epstopdf}
\usepackage{color}
\usepackage{verbatim} 
\usepackage{multirow}
\usepackage[abs]{overpic}
\usepackage{amsmath}

\newcommand{\PreserveBackslash}[1]{\let\temp=\\#1\let\\=\temp}
\newcolumntype{C}[1]{>{\PreserveBackslash\centering}p{#1}}
\newcolumntype{R}[1]{>{\PreserveBackslash\raggedleft}p{#1}}
\newcolumntype{L}[1]{>{\PreserveBackslash\raggedright}p{#1}}

\newcommand{\EE}{e^+e^-}

\newcommand{\psip}{\psi(3686)}

\newcommand{\jpsi}{J/\psi}
\newcommand{\too}{\rightarrow}

\begin{document}
\graphicspath{{figure/}}
\DeclareGraphicsExtensions{.eps,.png,.ps}

\title{\quad\\[0.0cm] \boldmath Study of electromagnetic Dalitz decays $\chi_{cJ} \too \mu^{+}\mu^{-}J/\psi$}

\author{
\begin{small}
    \begin{center}
      M.~Ablikim$^{1}$, M.~N.~Achasov$^{10,d}$, S. ~Ahmed$^{15}$, M.~Albrecht$^{4}$, M.~Alekseev$^{55A,55C}$, A.~Amoroso$^{55A,55C}$, F.~F.~An$^{1}$, Q.~An$^{52,42}$, Y.~Bai$^{41}$, O.~Bakina$^{27}$
% ...
      \\
      \vspace{0.2cm}
      (BESIII Collaboration)\\
      \vspace{0.2cm} {\it
        $^{1}$ Institute of High Energy Physics, Beijing 100049, People's Republic of China\\
% ...
$^{4}$ Bochum Ruhr-University, D-44780 Bochum, Germany\\
% ...
$^{10}$ G.I. Budker Institute of Nuclear Physics SB RAS (BINP), Novosibirsk 630090, Russia\\
% ...
$^{15}$ Helmholtz Institute Mainz, Johann-Joachim-Becher-Weg 45, D-55099 Mainz, Germany\\
% ...
$^{27}$ Joint Institute for Nuclear Research, 141980 Dubna, Moscow region, Russia\\
% ...
$^{41}$ Southeast University, Nanjing 211100, People's Republic of China\\
$^{42}$ State Key Laboratory of Particle Detection and Electronics, Beijing 100049, Hefei 230026, People's Republic of China\\
% ...
$^{52}$ University of Science and Technology of China, Hefei 230026, People's Republic of China\\
% ...
$^{55}$ (A)University of Turin, I-10125, Turin, Italy; (B)University of Eastern Piedmont, I-15121, Alessandria, Italy; (C)INFN, I-10125, Turin, Italy\\
% ...
$^{d}$ Also at the Novosibirsk State University, Novosibirsk, 630090, Russia\\
% ...
      }\end{center}
    \vspace{0.4cm}
    \end{small}
  }


\begin{abstract}
 Using $4.48 \times 10^{8}$ $\psip$ events collected with the BESIII detector, we search for the decays $\chi_{cJ} \too \mu^{+}\mu^{-}J/\psi$ through the radiative decays $\psip \too \gamma\chi_{cJ}$, where $J=0,1,2$. The decays $\chi_{c1,2} \too \mu^{+}\mu^{-}J/\psi$ are observed, and the corresponding branching fractions are measured to be $\mathcal{B}(\chi_{c1} \too \mu^{+}\mu^{-}J/\psi) = (2.51 \pm 0.18 \pm 0.20)\times10^{-4}$ and $\mathcal{B}(\chi_{c2} \too \mu^{+}\mu^{-}J/\psi) = (2.33 \pm 0.18 \pm 0.29)\times10^{-4}$, where the first uncertainty is statistical and the second one systematic. No significant $\chi_{c0} \too \mu^{+}\mu^{-}J/\psi$ decay is observed, and the upper limit on the branching fraction is determined to be $2.0\times10^{-5}$ at 90\% confidence level. Also, we present a study of di-muon invariant mass dependent transition form factor for the decays $\chi_{c1,2} \too \mu^{+}\mu^{-}J/\psi$.
\end{abstract}


\maketitle
\section{I. INTRODUCTION}
The electromagnetic (EM) Dalitz decays $M_1 \too M_2\ell^{+}\ell^{-}$ ($M$ for meson, $\ell=e$ or $\mu$) provide information on the internal structure of the mesons and the interactions of the mesons with the electromagnetic field~\cite{dalitz,theory,theory2,theory3}. Such decays are well studied in light-quark meson sector~\cite{pdg}, but very rare in charm sector, let alone in bottom sector. The $q$-dependent transition form factor (TFF), where $q$ is the invariant mass of the lepton pair, serves as a sensitive probe to the inner structure of the mesons involved, thus provides crucial tests to the theoretical models developed to describe the nature of the mesons, especially the charmonium-like states which manifested exotic properties compared with conventional charmonium states. One example is the $X(3872)$; while it is a candidate for the radial excitation of the $P$-wave charmonium state $\chi_{c1}$, it is also a good candidate of $D\bar{D}^{*}$ molecule. Precision measurement of its EM Dalitz decays and comparison with those of $\chi_{c1}$ decays and the relevant theoretical models may eventually reveal its nature.

The branching fractions for $\chi_{cJ} \too \mu^{+}\mu^{-}J/\psi$ are predicted in Ref.~\cite{eechicjtheory}  (Throughout this paper, $\chi_{cJ}$ refers to $\chi_{c0}$, $\chi_{c1}$, and $\chi_{c2}$), and it is demonstrated that the $\mu^{+}\mu^{-}$ decay channels are more suitable for the investigation of $\chi_{cJ} \too \gamma^* J/\psi$ decay vertices than $e^{+}e^{-}$ decay channels, which have been observed at BESIII~\cite{eechicj}. Very recently, LHCb reported the observation of $\chi_{c1,2} \too \mu^{+}\mu^{-}J/\psi$~\cite{mumuchicj} and measured the $\chi_{c1,2}$ resonance parameters. At BESIII, since the branching fractions of $\psi(3686) \too \gamma\chi_{cJ}$ can be calculated very precisely, we can measure the absolute branching fractions of $\chi_{cJ} \too \mu^{+}\mu^{-}J/\psi$. The branching fractions in theoretical calculations are related to the TFF, so the measurements can provide more constraints on theoretical calculations about TFF correction~\cite{eechicjtheory}.

In this work, we report the branching fraction measurements of $\chi_{cJ} \too \mu^{+}\mu^{-}J/\psi$ by analysing the cascade decay $\psip \too \gamma\chi_{cJ}$, $\chi_{cJ} \too \mu^{+}\mu^{-} J/\psi$. Here, the $J/\psi$ is reconstructed in its decay to an $e^{+}e^{-}$ or $\mu^+\mu^-$ pair. This analysis uses a data sample of $(4.481\pm0.029) \times 10^{8}$ $\psip$ events~\cite{totalnumber2} taken with the BESIII detector~\cite{Ablikim:2009aa} operating at BEPCII~\cite{bepcii} in 2009 and 2012. In addition, a data sample corresponding to an integrated luminosity of \text{$(44.49\pm0.02\pm0.44)$ pb$^{-1}$}, taken at $\sqrt{s} = 3.65$ GeV~\cite{data3650}, is used to estimate the background from continuum processes.

\section{II. BESIII DETECTOR AND MONTE CARLO SIMULATION}

The BESIII detector is a magnetic spectrometer~\cite{Ablikim:2009aa} located at the BEPCII~\cite{bepcii}. The
cylindrical core of the BESIII detector consists of a helium-based
 multilayer drift chamber (MDC), a plastic scintillator time-of-flight
system (TOF), and a CsI(Tl) electromagnetic calorimeter (EMC),
which are all enclosed in a superconducting solenoidal magnet
providing a 1.0~T magnetic field. The solenoid is supported by an
octagonal flux-return yoke with resistive plate counter muon
identifier modules interleaved with steel. The acceptance of
charged particles and photons is 93\% over $4\pi$ solid angle. The
charged-particle momentum resolution at $1~{\rm GeV}/c$ is
$0.5\%$, and the $dE/dx$ resolution is $6\%$ for the electrons
from Bhabha scattering. The EMC measures photon energies with a
resolution of $2.5\%$ ($5\%$) at $1$~GeV in the barrel (end cap)
region. The time resolution of the TOF barrel part is 68~ps, while
that of the end cap part is 110~ps.

Simulated samples produced with the {\sc geant4}-based~\cite{geant4} Monte Carlo (MC) package which
includes the geometric description of the BESIII detector and the
detector response, are used to determine the detection efficiency
and to estimate the backgrounds. The simulation includes the beam
energy spread and initial state radiation (ISR) in the $e^+e^-$
annihilations modelled with the generator {\sc kkmc}~\cite{ref:kkmc}.
The signal MC samples are generated using {\sc evtgen}~\cite{ref:evtgen} with a $q$-dependent decay amplitude based on the assumption of a point-like meson, as described in Refs.~\cite{eechicjtheory, generator}. The inclusive MC sample consists of the production of the
$\psi(3686)$ resonance, the ISR production of the $J/\psi$, and
the continuum processes incorporated in {\sc
kkmc}~\cite{ref:kkmc}. The known decay modes are modelled with {\sc
evtgen}~\cite{ref:evtgen} using branching fractions taken from the
Particle Data Group~\cite{pdg}, and the remaining unknown decays
from the charmonium states with {\sc
lundcharm}~\cite{ref:lundcharm}. The final state radiations (FSR)
from charged final state particles are incorporated with the {\sc
photos} package~\cite{photos}.

\section{III. EVENT SELECTION}

Candidate events are required to have four charged tracks, with zero net charge, and at least one photon. For each charged track, the distance of the closest approach to the interaction point (IP) is required to be smaller than \text{1 cm} on the radial direction and smaller than 10 cm along the beam axis. The polar angle ($\theta$) of the tracks must be within the fiducial volume of the MDC $(|\cos\theta|<0.93)$.

Photons are reconstructed from isolated showers in the EMC which are at least $20^\circ$ away from the nearest charged track. The photon energy is required to be at least 25 MeV in the barrel region $(|\cos\theta|<0.8)$ or 50 MeV in the endcap region $(0.86<|\cos\theta|<0.92)$. In order to suppress electronic noise and energy depositions which are unrelated to the event, the time after the collision at which the photon is recorded in the EMC is required to satisfy $0\leq t \leq 700$ ns.

According to the study of signal MC, the tracks with momentum larger than 1~GeV/$c$ are assumed to be leptons from $J/\psi$ decay, otherwise they are considered as muons from $\chi_{cJ}$ decay. The EMC deposited energy is used to separate electrons and muons from $J/\psi$, leptons from the $J/\psi$ decay with energy deposited in EMC larger than 1.0 GeV are identified as electrons, less than \text{0.3 GeV} as muons. The $\jpsi$ signal is selected by requiring the invariant mass of the lepton pair to be in the mass region [3.085, 3.110] GeV/$c^{2}$. A vertex fit is performed on the four charged tracks to restrict the tracks originated from the IP. In order to reduce backgrounds and improve the mass resolution, a four-constraint (4C) kinematic fit is performed by constraining the total four momentum to that of the initial beams. All the photons are looped with the four tracks in the kinematic fit and only those with a $\chi^2< 40$ are retained. If there is more than one photon candidate in an event, only the one with the least $\chi^{2}$ is retained for further analysis.

A study of the $\psip$ inclusive MC sample shows that, after applying the above selection criteria, the main backgrounds come from the four processes: Cat. I: $\psi(3686)\to\gamma\chi_{cJ}$, $\chi_{cJ}\to\gamma J/\psi$, $J/\psi\to \ell^{+}\ell^{-}$; Cat. II: $\psi(3686)\to\pi^0_{1}\pi^0_{2} J/\psi$, $\pi^0_{1}\to\gamma\gamma$, $\pi^0_{2}\to\gamma\gamma$, $J/\psi\to \ell^{+}\ell^{-}$; Cat. III: $\psi(3686)\to\pi^0_{1}\pi^0_{2} J/\psi$, $\pi^0_{1}\to\gamma\gamma$, $\pi^0_{2}\to\gamma\EE$, $J/\psi\to \ell^{+}\ell^{-}$ and Cat. IV: $\psi(3686)\to\eta J/\psi$, $\eta\to\gamma\mu^+\mu^-$, $J/\psi\to \ell^{+}\ell^{-}$.

\begin{figure}[htbp]
\begin{center}
\begin{overpic}[width=0.32\textwidth]{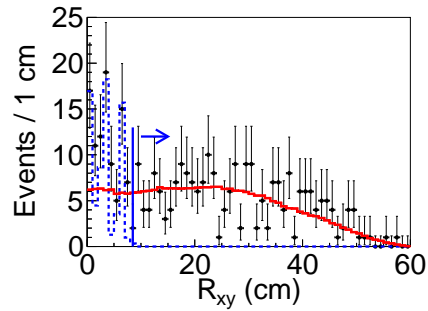}
\end{overpic}
\caption{Distribution of $R_{xy}$ for the decay $\chi_{c1} \too \mu^{+}\mu^{-}J/\psi$, where $R_{xy}$ is the distance from the reconstructed conversion vertex to the $z$ axis calculated from photo-conversion finder~\cite{conversion}. The points with error bars are data, the red histograms are for the signal MC simulations and the blue dotted lines are for the background MC simulations.}
\label{fig:conversion}
\end{center}
\end{figure}

To suppress the backgrounds from Cats. I and II, where one photon is converted into two electrons, a photon-conversion finder~\cite{conversion} is used to reconstruct the photon-conversion vertex. There are no additional requirements in photon-conversion finder. The distance from the reconstructed conversion vertex to the $z$ axis, $R_{xy}$, is used to distinguish the photon conversion background from the signal. Figure~\ref{fig:conversion} shows the $R_{xy}$ distribution of the decay $\chi_{c1} \too \mu^{+}\mu^{-}J/\psi$ as an example. By studying the MC samples of Cats. I and II, the peaks around $R_{xy}=3$ and $6$ cm match the positions of the beam pipe and the inner wall of MDC~\cite{Ablikim:2009aa}, respectively. For the background from Cat. III, it enhances around $R_{xy}=0$ cm. In order to remove all these backgrounds, a requirement $R_{xy} > 8.5$ cm is applied. For the signal events, the reconstructed $R_{xy}$ is almost proportional to the opening angle of the two $\mu$ tracks, so if the angle of the two tracks is large, the variable $R_{xy}$ is also large.

To remove the background from Cat. IV, which has the same final state as the signal event, a requirement $M(\gamma \mu^{+}\mu^{-}) < 0.535$ or $> 0.560$ GeV/$c^2$ is applied, where $M(\gamma \mu^{+}\mu^{-})$ is the invariant mass of $\gamma \mu^{+}\mu^{-}$. The requirement removes almost all the Cat. IV background events (it accounts for about $60\%$ of the remaining background, is about 140 events), with an efficiency loss of about $15\%$ for signal events.

After applying all the above criteria for $\psip$ inclusive MC sample, which does not include the signal processes, only a few events are left, and the overall contribution from $\psip$ decays in the $M(\mu^{+}\mu^{-} J/\psi)$ distribution is found to be smooth. Here $M(\mu^+\mu^- J/\psi)=M(\mu^+\mu^- \ell^{+}\ell^{-})-M(\ell^{+}\ell^{-})+m(J/\psi)$ is used to reduce the resolution effect of the lepton pairs, and $m(J/\psi)$ is the nominal mass of $J/\psi$~\cite{pdg}. The continuum background is studied by using the data collected at $\sqrt{s}=3.65$ GeV, and the contribution is found to be negligible.

\begin{table*}[htbp]
\caption{Signal yields, detection efficiency, branching fraction (or upper limit at $90\%$ C.L.) and ratio of the branching fractions for each decay channel. Here the first uncertainty is statistical and the second systematic. }
\label{tab:branching}
\setlength{\tabcolsep}{4mm}{
\begin{tabular}{ c  c  c  c  c }
  \hline
  \hline
  Decay mode & Yields & Efficiency ($\%$) & Branching fraction & $\frac{\mathcal{B}(\chi_{cJ} \too \mu^{+}\mu^{-}J/\psi)}{\mathcal{B}(\chi_{cJ} \too e^+e^-J/\psi)}$ \\
  \hline
  $\chi_{c0} \too \mu^{+}\mu^{-} J/\psi$ & $<9.5$   & 9.40 & $<2.0\times10^{-5}$ & $<0.14$ \\
  $\chi_{c1} \too \mu^{+}\mu^{-} J/\psi$ & $221.9\pm15.3$ & 16.94 & $(2.51 \pm 0.18 \pm 0.20)\times10^{-4}$ & $(6.73 \pm 0.51 \pm 0.50)\times10^{-2}$ \\
  $\chi_{c2} \too \mu^{+}\mu^{-} J/\psi$ & $218.9\pm16.1$ & 18.42 & $(2.33 \pm 0.18 \pm 0.29)\times10^{-4}$ & $(9.40 \pm 0.79 \pm 1.15)\times10^{-2}$ \\
  \hline
  \hline
\end{tabular}}
\end{table*}

\begin{table*}[htbp]
\caption{Signal yields, measured branching fraction $\mathcal{B}$, QED predicted branching fraction $\mathcal{B}_{\text{QED}}$~\cite{eechicjtheory} and TFF $|F(q)|^2$ for the decays $\chi_{c1,2}\too\mu^+\mu^-J/\psi$ in each bin. Here the first uncertainty is statistical and the second systematic. }
\label{tab:tff}
\setlength{\tabcolsep}{4mm}{
\begin{tabular}{c  c  c  c c c  }
  \hline
  \hline
  Channel & $q$ (GeV/$c^2$) & Yields & $\mathcal{B}$ ($10^{-5}$) & $\mathcal{B}_{\text{QED}}$~\cite{eechicjtheory} ($10^{-5}$) & $|F(q)|^2$ \\
  \hline
  $\chi_{c1}\too\mu^+\mu^-J/\psi$& $[2m_{\mu}, 0.25]$ & $26.9\pm5.4$   & $4.32\pm0.87\pm0.35$ & $3.81\pm0.11$ & $1.13\pm0.23\pm0.10$ \\
  & $[0.25, 0.30]$ & $74.4\pm8.9$ & $8.87\pm1.06\pm0.71$ & $5.91\pm0.17$ & $1.50\pm0.18\pm0.13$ \\
  & $[0.30, 0.35]$ & $63.4\pm8.3$ & $6.51\pm0.85\pm0.52$ & $4.64\pm0.14$ & $1.40\pm0.19\pm0.12$ \\
  & $[0.35, 0.40]$ & $59.5\pm7.9$ & $5.17\pm0.69\pm0.42$ & $2.83\pm0.08$ & $1.83\pm0.25\pm0.16$ \\
  \hline
  $\chi_{c2}\too\mu^+\mu^-J/\psi$& $[2m_{\mu}, 0.25]$ & $29.1\pm5.9$   & $4.20\pm0.85\pm0.52$ & $2.20\pm0.06$  & $1.91\pm0.39\pm0.24$ \\
  & $[0.25, 0.30]$ & $50.7\pm7.8$ & $5.32\pm0.82\pm0.66$ & $3.51\pm0.09$ & $1.52\pm0.24\pm0.19$ \\
  & $[0.30, 0.35]$ & $47.4\pm7.7$ & $5.02\pm0.82\pm0.62$ & $2.93\pm0.08$ & $1.71\pm0.28\pm0.22$ \\
  & $[0.35, 0.40]$ & $56.9\pm7.9$ & $5.61\pm0.78\pm0.70$ & $2.16\pm0.06$ & $2.60\pm0.37\pm0.33$ \\
  & $[0.40, 0.45]$ & $38.3\pm6.8$ & $3.45\pm0.62\pm0.43$ & $1.25\pm0.03$ & $2.76\pm0.50\pm0.35$ \\
  \hline
  \hline
\end{tabular}}
\end{table*}

\section{IV. BRANCHING FRACTION MEASUREMENT}

\begin{figure}[htbp]
\begin{center}
\begin{overpic}[width=0.36\textwidth]{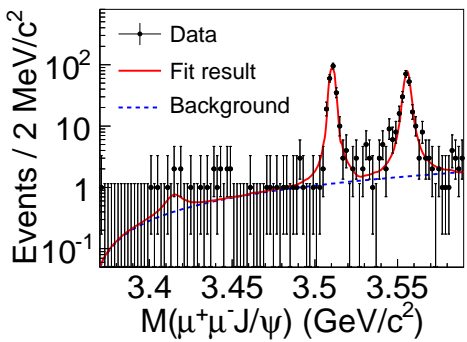}
\end{overpic}
\caption{ Distribution of $M(\mu^{+}\mu^{-} J/\psi)$ in data (dots with error bars). The solid curve is the overall fit result, the dashed curve is for background contribution.}
\label{fig:fitresult}
\end{center}
\end{figure}

Figure~\ref{fig:fitresult} shows the $M(\mu^{+}\mu^{-} J/\psi)$ distribution for selected events from data. Clear enhancements at the masses of $\chi_{c1,2}$ are seen, corresponding to the decays $\chi_{c1,2} \too \mu^{+}\mu^{-}J/\psi$, while no significant signals for the $\chi_{c0} \too \mu^{+}\mu^{-}J/\psi$ decay are found. An unbinned maximum likelihood fit is performed to the $M(\mu^{+}\mu^{-} J/\psi)$ distribution to extract the signal yields. We use the MC-determined shapes to describe the $\chi_{cJ}$ signals, where the magnitudes are free parameters. The background is described by a linear function with the number of events as free parameter. The fit result is shown in Fig.~\ref{fig:fitresult} and the corresponding signal yields are summarized in Table~\ref{tab:branching}. The significances for $\chi_{c1,2}$ are larger than 10$\sigma$ by comparing the likelihood values for the fits with or without $\chi_{c1,2}$ signals and taking the change of the number of degrees-of-freedom into account. Since no significant signal is observed for $\chi_{c0} \too \mu^{+}\mu^{-}J/\psi$ decay, we give the upper limit at $90\%$ confidence level (C.L.) using Bayesian method. With the fit function described before, we scan the number of $\chi_{c0}$ signal yield to obtain the likelihood distribution, and smear it with the systematic uncertainty. The upper limit of the number of $\chi_{c0}$ signal yield $N^{up}_{\chi_{c0}}$ at $90\%$ C.L. is obtained via $\int_{0}^{N^{up}_{\chi_{c0}}}F(x)dx/\int_{0}^{\infty}F(x)dx=0.90$, where $F(x)$ is the probability density function of the likelihood distribution.

The branching fractions $\mathcal{B}(\chi_{cJ} \too \mu^{+}\mu^{-}J/\psi)$ are calculated according to
\begin{footnotesize}
\begin{equation}
    \mathcal{B}(\chi_{cJ} \too \mu^{+}\mu^{-}J/\psi) = \frac{N}{N_{\psi(3686)}\cdot{\mathcal{B}_{\text{rad}}}\cdot{\mathcal{B}_{J/\psi \too l^{+}l^{-}}}\cdot{\epsilon}},
\end{equation}
\end{footnotesize}where $N$ is the signal yields obtained from the fit, $N_{\psip}$ is the number of $\psip$ events~\cite{totalnumber2}, $\epsilon$ is the average selection efficiency of the decays $J/\psi \too e^{+}e^{-}$ and $J/\psi \too \mu^{+}\mu^{-}$ determined from the signal MC samples, $\mathcal{B}_{\mathrm{rad}}$ is the branching fraction of the radiative transitions $\psip \too \gamma\chi_{cJ}$, and $\mathcal{B}_{J/\psi \too \ell^{+}\ell^{-}}$ is the sum of branching fractions of $J/\psi \too e^{+}e^{-}$ and $J/\psi \too \mu^{+}\mu^{-}$. All the branching fractions used are taken from Ref.~\cite{pdg}. The results of $\chi_{cJ} \too \mu^{+}\mu^{-}J/\psi$ are listed in Table~\ref{tab:branching}.

\section{V. Transition form factor MEASUREMENT}

\begin{figure}[htbp]
\begin{center}
\begin{overpic}[width=0.23\textwidth]{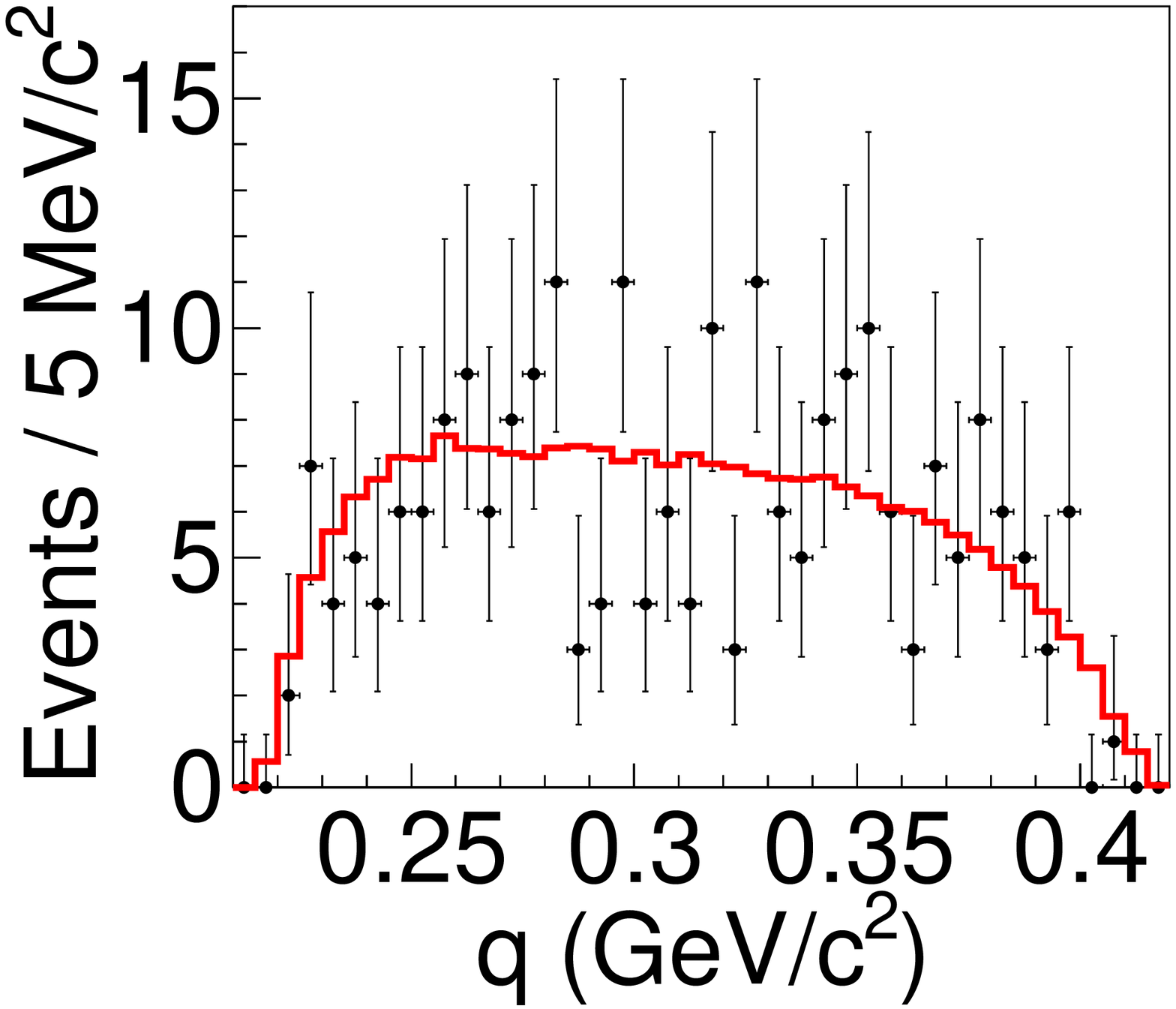}
\put(95,80){(a)}
\end{overpic}
\begin{overpic}[width=0.23\textwidth]{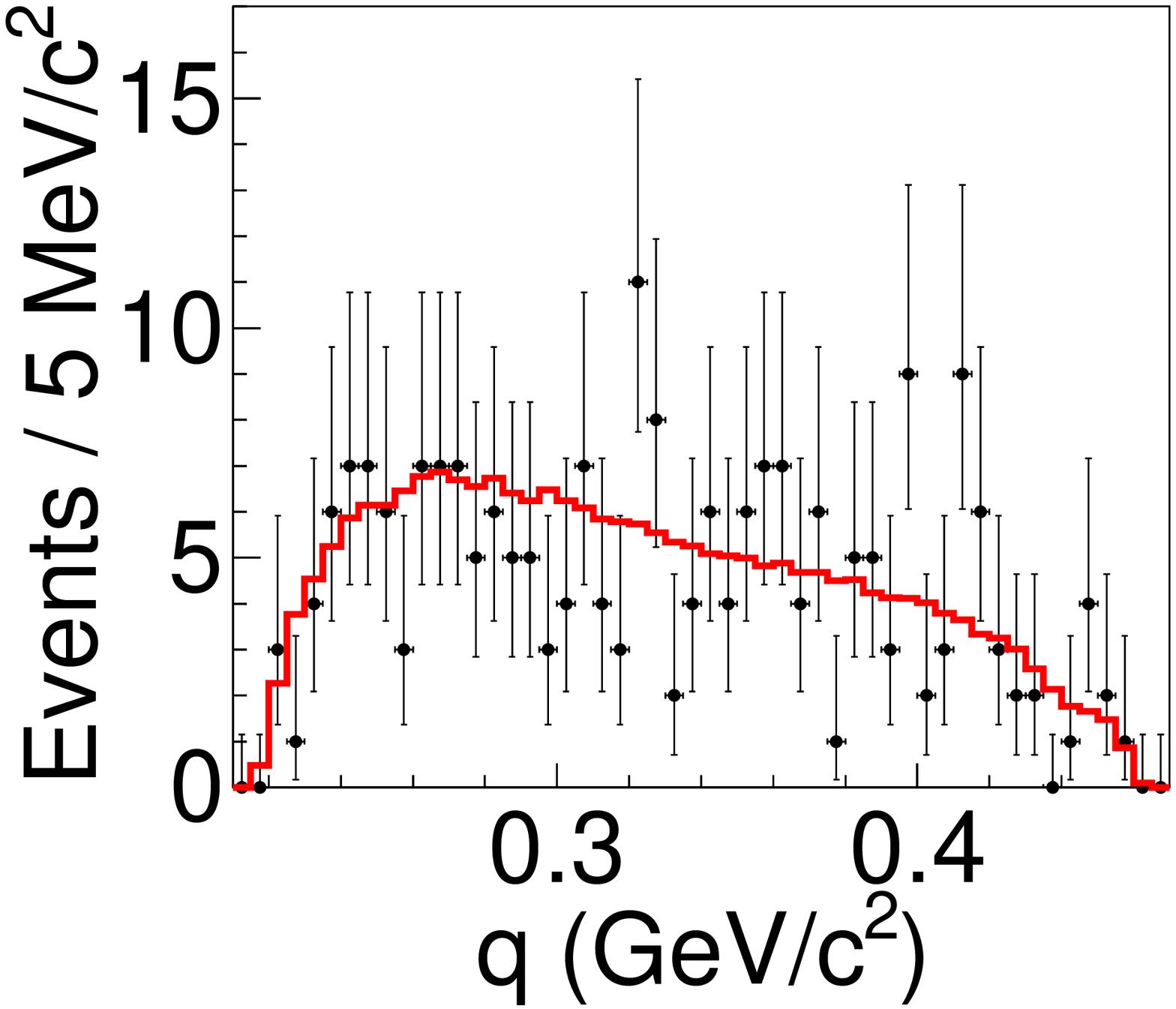}
\put(95,80){(b)}
\end{overpic}
\caption{Comparison of $q$ distributions between data and MC simulation. The distributions are not efficiency corrected for the decays $\chi_{c1} \too \mu^{+}\mu^{-}J/\psi$ (a) and $\chi_{c2} \too \mu^{+}\mu^{-}J/\psi$ (b). The points with error bars are data and the red histograms are for the signal MC simulations. The MC distributions are normalized by the total number of events for data.}
\label{fig:eemass}
\end{center}
\end{figure}

Figure~\ref{fig:eemass} shows comparisons of the observed $q$ distributions without efficiency correction in data and MC simulation for the decays $\chi_{c1,2} \too \mu^{+}\mu^{-}J/\psi$, where the $\chi_{c1}$ and $\chi_{c2}$ signals are extracted by requiring a mass within [3.500, 3.520] and [3.545, 3.565] GeV/$c^2$, respectively; with these criteria the backgrounds are expected to be about 5$\%$. The data are in reasonable agreement with the MC simulation generated by using the point-like  model described in Refs.~\cite{eechicjtheory, generator}.

\begin{figure}[htbp]
\begin{center}
\begin{overpic}[width=0.23\textwidth]{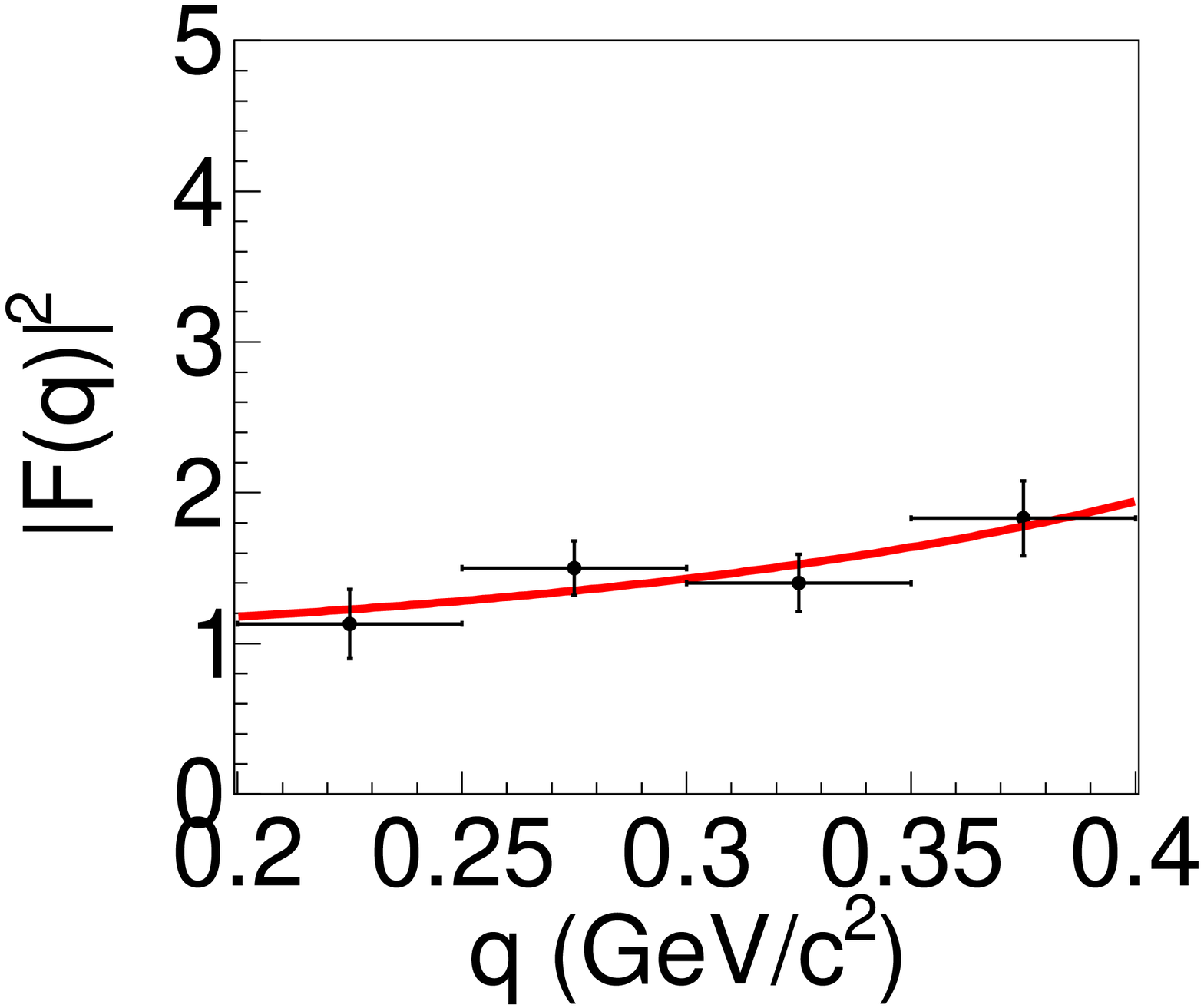}
\put(95,80){(a)}
\end{overpic}
\begin{overpic}[width=0.23\textwidth]{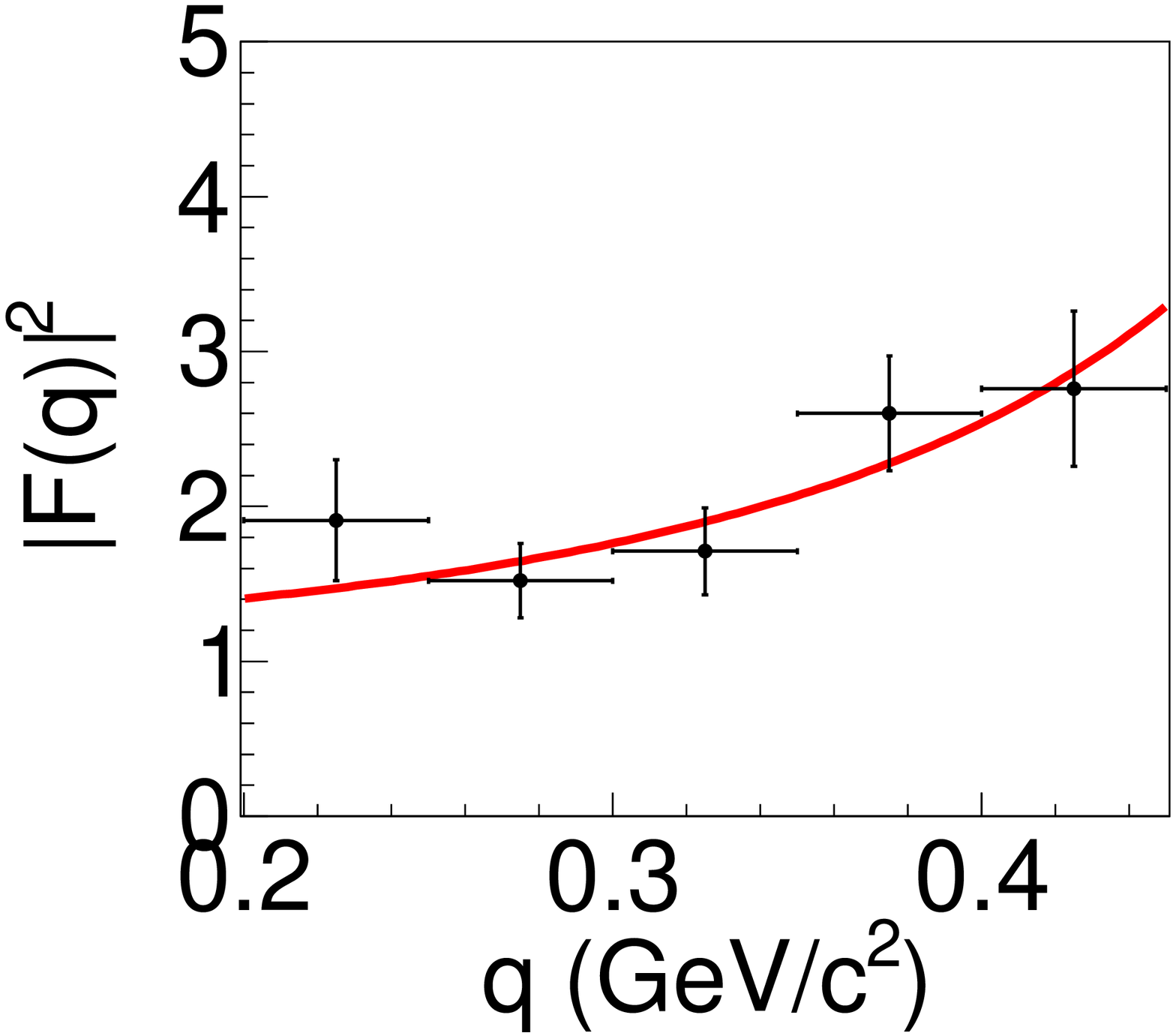}
\put(95,80){(b)}
\end{overpic}
\caption{TFF distributions for the decays $\chi_{c1} \too \mu^{+}\mu^{-}J/\psi$ (a) and $\chi_{c2} \too \mu^{+}\mu^{-}J/\psi$ (b). The solid curves are the fit results.}
\label{fig:tff}
\end{center}
\end{figure}

To measure the TFF, the $q$ distributions in the decays $\chi_{c1,2} \too \mu^{+}\mu^{-}J/\psi$ are divided into 4 and 5 regions, respectively. The bin-by-bin signal yields and corresponding branching fractions are listed in Table~\ref{tab:tff}. The quantum electrodynamics (QED) predicted branching fraction results of $\chi_{c1,2} \too \mu^{+}\mu^{-}J/\psi$ are obtained from Eq.(2) in Ref.~\cite{eechicjtheory}, and the uncertainty is from the branching fractions of $\chi_{c1,2} \too \gamma J/\psi$. The TFFs are the ratios of measured branching fractions and QED predicted branching fractions in each bin, which are also listed in Table~\ref{tab:tff}. Figure~\ref{fig:tff} shows the TFF distributions for the decays $\chi_{c1,2} \too \mu^{+}\mu^{-}J/\psi$. If we use the parametrization $F(q)=\frac{1}{1-q^2/\Lambda^2}$~\cite{theory2} to fit TFF distributions, the fit results are also shown in Fig.~\ref{fig:tff}. The $\Lambda$ values for the decays $\chi_{c1,2} \too \mu^{+}\mu^{-}J/\psi$ are $\Lambda_{\chi_{c1}}=(0.76\pm0.18)$ GeV/$c^2$ and $\Lambda_{\chi_{c2}}=(0.71\pm0.10)$ GeV/$c^2$.

\section{VI. SYSTEMATIC UNCERTAINTY}

The systematic uncertainties for the branching fraction measurement arise from the following sources: track reconstruction, photon detection, kinematic fit, $J/\psi$ mass criteria, $M(\gamma \mu^{+}\mu^{-})$ requirement, $R_{xy}$ requirement, fit procedure, angular distribution, number of $\psip$ events, and the branching fractions of the cascade decays. All uncertainties are discussed in detail below.

The difference between data and MC simulation on the tracking efficiency of high momentum tracks is estimated to be $1\%$~\cite{trackeff} using control sample $\psi(3686) \too \pi^+\pi^-J/\psi$, $J/\psi \too \ell^+\ell^-$. To study the difference on the low momentum muon tracking efficiency between data and MC simulation, we select a sample of $\psi(3686) \too \pi^+\pi^-J/\psi$, $J/\psi \too \mu^+\mu^-\gamma$. The weighted difference between data and MC simulation is about $4\%$ for the low momentum $\mu^+\mu^-$ pair. We also checked $\rm{cos}\theta$ dependence of low momentum tracking efficiency using control sample $J/\psi \too p\bar{p}\pi^+\pi^-$. The $\pi$ tracking efficiency is $\rm{cos}\theta$ dependent, and we use these results to correct the efficiency for $\mu^+\mu^-$ pair, while the weighted difference between data and MC simulation is also about $4\%$. Totally, a $6\%$ systematic uncertainty on tracking efficiency is attributed to all channels. The uncertainty on the photon detection efficiency is derived from a control sample of $J/\psi \too \rho^{0}\pi^{0}$ and is $1.0\%$ per photon~\cite{rhopi}.

In the 4C kinematic fit, the helix parameters of charged tracks are corrected to reduce the discrepancy between data and MC simulation as described in Ref.~\cite{helix}. The correction factors are obtained by studying a control sample of $\psip \too \pi^{+}\pi^{-}J/\psi, J/\psi \too \ell^{+}\ell^{-}$. To determine the systematic uncertainty from this source, we determine the efficiencies from the MC samples without the helix correction; the resulting differences with respect to the nominal values are taken as systematic uncertainties.

The uncertainty associated with the $J/\psi$ mass requirement is 1.0\%, which is determined by studying a control sample of $\psip \too \eta J/\psi, \eta \too \gamma\gamma$ (where one $\gamma$ undergoes conversion to an $e^+e^-$ pair) or $\eta \too \gamma e^{+}e^{-}$ decays. The systematic uncertainty related to the $M(\gamma \mu^{+}\mu^{-})$ requirement is studied by removing the requirement and then repeat the analysis to get the result. The difference from the nominal result is taken as systematic uncertainty from this source. Likewise to estimate the systematic uncertainty from $R_{xy}$ requirement, we also remove the requirement to get the result and the difference is taken as systematic uncertainty. Due to the absence of $\chi_{c0}$ signal, the uncertainties for $\chi_{c0}$ channel on $M(\gamma \mu^{+}\mu^{-})$ requirement and $R_{xy}$ requirement are taken from the larger one in the $\chi_{c1}$ and $\chi_{c2}$ channels.

The sources of uncertainty in the fit procedure include the fit range, the signal shape, and the background shape. The uncertainty related to the fit range is obtained by varying the limits of the fit range by $\pm$5 MeV/$c^{2}$. The largest difference in the signal yields with respect to the nominal values is taken as systematic uncertainty. In the nominal fit, the signal shapes are described with the MC simulated signal shapes. An alternative fit is performed with the signal MC simulated shapes convolved with a Gaussian function. The resulting change in the signal yields is taken as systematic uncertainty. The uncertainty associated with the background shape is estimated by an alternative fit replacing the first order polynomial function with a second order polynomial function. The change in the signal yields is taken as systematic uncertainty. About the uncertainty from fit procedure for $\chi_{c0}$ channel, we try to use different combinations of fit range, signal shape and background shape to get the upper limits, and choose the largest one as nominal upper limit.

The helicity angle distribution $1+\alpha\cdot\rm{cos}^2\theta$ of the $\mu^{+}\mu^{-}$ pair in $\chi_{cJ}$ rest frame may affect the detection efficiency, where $\alpha$ is angular distribution parameter. We fix the angular distribution of $\mu^{+}\mu^{-}$ pair in $\chi_{c1,2} \too \mu^{+}\mu^{-}J/\psi$ at the measurements from processes $\chi_{c1,2} \too \EE J/\psi$~\cite{eechicj}, $\alpha_{\chi_{c1}}=0.0\pm0.2$ and $\alpha_{\chi_{c2}}=0.5\pm0.2$, and vary 1$\sigma$ to get the systematic uncertainty. For the decay $\chi_{c0} \too \mu^{+}\mu^{-}J/\psi$, the angular distribution of $\mu^{+}\mu^{-}$ pair is set to be flat, and varied to $1\pm$cos$^2\theta$ to get the systematic uncertainty.

The number of $\psip$ events is measured with an uncertainty of 0.7\% by using the inclusive hadronic events~\cite{totalnumber2}. The uncertainties of the branching fractions in the cascade decays are taken from Ref.~\cite{pdg}.

Table~\ref{tab:sumerror} summarizes all individual systematic uncertainties, and the overall uncertainties are the quadrature sums of the individual ones, assuming they are independent. The overall uncertainties are also taken as the systematic uncertainties of the branching fraction measurements for the decays $\chi_{c1,2}\too\mu^+\mu^-J/\psi$ in each bin in Table~\ref{tab:tff}.

\begin{table}[htbp]
\caption{Summary of systematic uncertainties (in $\%$). Dash means that the results are not applicable.}
\label{tab:sumerror}
\begin{tabular}{ cccc }
  \hline
  \hline
   &\multicolumn{3}{c}{~$\chi_{cJ} \too \mu^{+}\mu^{-}J/\psi$~~}  \\
   &~~~$\chi_{c0}$~~ &~~$\chi_{c1}$~~&~~$\chi_{c2}$~~ \\
  \hline
  Tracking                    & 6.0& 6.0& 6.0  \\

  Photon                      & 1.0& 1.0&  1.0  \\

  Kinematic fit               & 2.6& 2.5& 2.5  \\

  $J/\psi$ mass window        & 1.0& 1.0& 1.0  \\

  $M(\gamma \mu^{+}\mu^{-})$ requirement      & 6.0& 1.8& 6.0  \\

  $R_{xy}$ requirement & 7.6& 2.7& 7.6  \\

  Fit range                  & $-$& 0.5& 2.1  \\

  Signal shape                & $-$& 0.1& 0.4  \\

  Background shape            & $-$& 0.2& 2.3  \\

  Angular distribution        & 2.7& 1.1& 0.1  \\

  Number of $\psip$           & 0.7& 0.7& 0.7  \\

  Branching fractions         & 2.1& 2.5& 2.1  \\

  \hline

  sum                         & 12.3& 8.0& 12.4  \\
  \hline
  \hline
\end{tabular}
\end{table}

\section{VII. RESULTS AND DISCUSSION}

In summary, we observe the decays $\chi_{c1,2} \too \mu^{+}\mu^{-}J/\psi$ through the radiative transitions $\psip \too \gamma\chi_{cJ}$. The corresponding branching fractions are measured for the first time to be $\mathcal{B}(\chi_{c1} \too \mu^{+}\mu^{-}J/\psi) = (2.51 \pm 0.18 \pm 0.20)\times10^{-4}$ and $\mathcal{B}(\chi_{c2} \too \mu^{+}\mu^{-}J/\psi) = (2.33 \pm 0.18 \pm 0.29)\times10^{-4}$, where the first uncertainty is statistical and the second systematic. We do not observe significant $\chi_{c0} \too \mu^{+}\mu^{-}J/\psi$ events, and an upper limit at 90\% C.L. on the branching fraction is set to be $\mathcal{B}(\chi_{c0} \too \mu^{+}\mu^{-}J/\psi) < 2.0\times10^{-5}$. The ratios of branching fractions $\frac{\mathcal{B}(\chi_{cJ} \too \mu^{+}\mu^{-}J/\psi)}{\mathcal{B}(\chi_{cJ} \too e^+e^-J/\psi)}$ are also obtained by incorporating the BESIII measurements of the branching fractions $\mathcal{B}(\chi_{cJ} \too e^+e^-J/\psi)$ in Ref.~\cite{eechicj}, as listed in Table~\ref{tab:branching}. The common systematic uncertainties related to efficiency and branching fractions cancel in the calculation. From the measured TFF distributions, the $|F(q)|^2$ values deviate from one significantly. This indicates that the TFF should be considered in the branching fraction calculation. If we use the parametrization $F(q)=\frac{1}{1-q^2/\Lambda^2}$ to parameterize TFF with $\Lambda=m_{\rho}=0.77$ GeV/$c^2$, the calculated branching fractions~\cite{eechicjtheory} for $\chi_{cJ} \too \mu^{+}\mu^{-}J/\psi$ agree well with the measured results.

\section{ACKNOWLEDGMENTS}

The BESIII collaboration thanks the staff of BEPCII and the IHEP computing center for their strong support. This work is supported in part by National Key Basic Research Program of China under Contract No. 2015CB856700; National Natural Science Foundation of China (NSFC) under Contracts Nos. 11847028, 11335008, 11425524, 11625523, 11635010, 11735014, 11505034, 11521505, 11575198, U1732105; the Chinese Academy of Sciences (CAS) Large-Scale Scientific Facility Program; the CAS Center for Excellence in Particle Physics (CCEPP); Joint Large-Scale Scientific Facility Funds of the NSFC and CAS under Contracts Nos. U1532257, U1532258, U1732263; CAS Key Research Program of Frontier Sciences under Contracts Nos. QYZDJ-SSW-SLH003, QYZDJ-SSW-SLH040; 100 Talents Program of CAS; INPAC and Shanghai Key Laboratory for Particle Physics and Cosmology; Foundation of Henan Educational Committee (No. 19A140015); Nanhu Scholars Program for Young Scholars of Xinyang Normal University; German Research Foundation DFG under Contract No. Collaborative Research Center CRC 1044; Istituto Nazionale di Fisica Nucleare, Italy; Koninklijke Nederlandse Akademie van Wetenschappen (KNAW) under Contract No. 530-4CDP03; Ministry of Development of Turkey under Contract No. DPT2006K-120470; National Science and Technology fund; The Swedish Research Council; U. S. Department of Energy under Contracts Nos. DE-FG02-05ER41374, DE-SC-0010118, DE-SC-0010504, DE-SC-0012069; University of Groningen (RuG) and the Helmholtzzentrum fuer Schwerionenforschung GmbH (GSI), Darmstadt.

\end{document}